\documentclass{article}
\usepackage{setspace}
\usepackage[left=1in, right=1in, top=1.25in, bottom=1.25in]{geometry}
\usepackage[utf8]{inputenc}
\usepackage{graphicx,subcaption}
\usepackage{amsmath}
\usepackage{booktabs}
\usepackage{multirow}
\usepackage{longtable}
\usepackage{hyperref}
\usepackage[nottoc]{tocbibind}
\usepackage{arydshln}
\usepackage[table]{xcolor}
\usepackage{pdfpages}
\usepackage{comment}
\usepackage{adjustbox}
\usepackage{wrapfig}
\usepackage{floatrow}
\newfloatcommand{capbtabbox}{table}[][\FBwidth]

\usepackage[%
  autocite    = superscript,
  backend     = bibtex,
  sortcites   = none,
  citestyle   = numeric-comp,
  bibstyle    = numeric,
  sorting     = none,
  maxbibnames = 99,
  ]{biblatex}

\title{\textbf{Automatic Stroke Classification of Tabla Accompaniment in Hindustani Vocal Concert Audio}}
\date{}

\author{\normalsize{\textit{To appear in the JOURNAL OF ACOUSTICAL SOCIETY OF INDIA, April 2021}} \\ \\ Rohit M. A. and Preeti Rao\\ Department of Electrical Engineering \\ Indian Institute of Technology Bombay, Mumbai, India \\ \{rohitma, prao\}@ee.iitb.ac.in}

\begin{document}

\maketitle

\section*{ABSTRACT}
The tabla is a unique percussion instrument due to the combined harmonic and percussive nature of its timbre, and the contrasting harmonic frequency ranges of its two drums. This allows a tabla player to uniquely emphasize parts of the rhythmic cycle (\textit{theka}) in order to mark the salient positions. An analysis of the loudness dynamics and timing deviations at various cycle positions is an important part of musicological studies on the expressivity in tabla accompaniment. To achieve this at a corpus-level, and not restrict it to the few recordings that manual annotation can afford, it is helpful to have access to an automatic tabla transcription system. Although a few systems have been built by training models on labeled tabla strokes, the achieved accuracy does not necessarily carry over to unseen instruments. In this article, we report our work towards building an instrument-independent stroke classification system for accompaniment tabla based on the more easily available tabla solo audio tracks. We present acoustic features that capture the distinctive characteristics of tabla strokes and build an automatic system to predict the label as one of a reduced, but musicologically motivated, target set of four stroke categories. 
To address the lack of sufficient labeled training data, we turn to common data augmentation methods and find the use of pitch-shifting based augmentation to be most promising. We then analyse the important features and highlight the problem of their instrument-dependence while motivating the use of more task-specific data augmentation strategies to improve the diversity of training data. \\

\section{Introduction}
Hindustani classical music is a genre of art music in India encompassing several sub-genres and forms of presentation. \textit{Khyal}, one of the more modern forms, with roots in Dhrupad, consists of vocals as the lead, the harmonium (or sometimes the saarangi) for melodic accompaniment, the tabla for percussion accompaniment, and the tanpura in the background for the harmonic drone \cite{wade1979musicindia}{}. The performance opens with a short unmetered raga improvisation (\textit{alap}) by the vocalist, performed to a mild accompaniment by the harmonium, and is followed by the metered piece - the bandish, where the tabla also joins in. For the most part, the melodic accompaniment shadows the vocals while the tabla plays a fairly fixed cyclic pattern of beats called the `theka’ to provide the rhythm.\\

The tabla is a pitched percussive membranophone, consisting of a pair of drums called the \textit{bayan} and \textit{dayan}, or \textit{dagga} and \textit{tabla}, respectively. A unique property of these drums that sets them apart from common percussion instruments is their ability to produce ringing, pitched sounds. This is facilitated by the additional metallic layer at the center of each drum's membrane. When struck freely, the bayan, bigger in size, produces bass frequency sounds (F0\footnote{Fundamental frequency} $\in$ 80-100 Hz), while the dayan produces treble frequency sounds (F0 $\in$ 200-400 Hz). We therefore refer to these drums as bass and treble in the remainder of this article. The treble drum, tuned to the tonic note of the singing in a performance, is more harmonic than the bass drum which evokes only a faint sense of pitch.\\

Due to the predominantly improvisatory nature of Hindustani music, and a lack of precise performance-oriented theory, a greater emphasis has come to be placed on the empirical analysis of performed concerts. While traditional musicological studies would often be restricted to a few chosen concerts, the advancement of computational methods has made it possible to analyse large corpora. The focus of some of these studies has been to quantify aspects of rhythm and tempo in khyal and instrumental music concerts from the perspective of the percussion accompaniment, such as - stroke loudness dynamics and beat-level timing deviations in the tabla playing \cite{ajay2017rhythm}{}, typical tempo ranges in khyal performance and the variations in theka based on the chosen tempo \cite{clayton2020metre}{}, and the possible interaction of this with the accompanied lead vocals or melodic instrument. There is thus a large scope to help further such research by making use of automatic tools for accurate tabla transcription to aid large-scale analyses. \\

Supervised methods in machine learning achieve great success when provided sufficient labeled data that is matched to the target domain. In the present context, our target data, viz. tabla accompaniment from vocal concerts is available as such only in specially created multi-track recordings where each instrument is recorded with a separate microphone with sufficient physical separation between the artists. Further, although musical source separation is a rapidly developing area, there is no model currently available for the Hindustani vocal concert context. So we develop our models on the more amply available tabla solo, hoping to extend the trained models to accompaniment tabla obtained from multi-track recordings or possibly future source separation techniques. 
In the next section, we review the previous work in tabla stroke classification. We then present our problem and dataset, followed by a discussion of our method and a critical evaluation of its performance on our test data.  
\\

\section{Background}\label{sec:background}
The approach to transcribing tabla most commonly found in literature is one where segments of individual strokes are first represented using low-level audio features and then classified into one of a set of 10 - 15 bols \cite{gillet2003tabla, chordia2004tabla, subodh2016tabla, subodh2018tabla, sarkar2018tabla, shete2020tabla}{}. The stroke segments are obtained either by extracting them from a sequence of tabla strokes using a separate segmentation method \cite{gillet2003tabla,chordia2004tabla,sarkar2018tabla}{}, or by building a dataset of individual strokes recorded in isolation \cite{subodh2016tabla,shete2020tabla}{}. The various systems reported so far can be compared via the following aspects - the set of features used to characterize strokes, the models used for classification, the nature and diversity of the dataset, and the evaluation methods.\\

There is not a significant difference in the choice of features, with a majority of the previous methods employing a similar set of somewhat generic timbral descriptors such as Mel-Frequency Cepstral Coefficients (MFCC), spectral distribution features like the spectral centroid, skewness, and kurtosis, Linear Prediction Coefficients (LPC), and temporal features like the zero crossing rate, temporal centroid, and attack time. A dimensionality reduction step has been additionally used to select the most important subset of these features \cite{chordia2004tabla,subodh2016tabla}{}. However, while it would appear, from the fairly high classification accuracies reported, that these common low-level timbre descriptors are well-suited to the task, it is important to note that these descriptors tend to also depend on the physical characteristics of the tabla and differences in stroke articulation \cite{patranabis2015harmonic}{}. Therefore, along with analysing the quality of features based on the resulting separability of strokes \cite{subodh2018tabla}{}, it is also important to consider the instrument-independence of these features.\\

For the task of classification, a number of common models have been used, ranging from simple decision trees, k-nearest neighbours, and single layer feed-forward neural networks, to the slightly more complex random forests, probabilistic neural networks, and hidden markov models (HMM). The use of an HMM, which models temporal context information, seems to have helped in cases where ambiguous stroke labels were present in the dataset \cite{gillet2003tabla,gupta2015tabla}{}. Such ambiguities arise commonly in recordings of tabla compositions, where similar sounding strokes sometimes have different labels and vice-versa. In other cases, where the labels were based more on the stroke acoustics, or where the dataset consisted of isolated strokes, high classification accuracies have been achieved even without modelling the context.\\

The task of transcribing strokes drawn from continuous playing warrants a consideration of a few additional points over isolated stroke recordings. One, regarding the use of language models, is whether temporal context information learnt from solo compositions would be appropriate for tabla accompaniment transcription. While training language models on the basic thekas is a possibility, this would not capture all of the expressivity observed in concerts. An additional challenge is the overlap of harmonics from a previous stroke with subsequent ones \cite{chordia2008tablagyan}{}. For instance, a damped stroke like `ke' (played on the bass drum) could be mistaken for a stroke like `Na' with a sustained sound (produced on the treble drum) occurring just before it, because of the harmonics from `Na' that continue to sound during `ke'. Hence, a system trained on isolated strokes is likely to be of limited value in analysing tabla playing in concert audios.\\

The evaluation method is important not only in assessing the classification accuracy of a proposed method but more importantly, its generalisation to unseen instruments. Due to the drastic timbre differences between tabla sets arising from diverse physical properties and tuning, building a robust classifier that works equally well on a variety of instruments has been found to be a difficult task \cite{chordia2004tabla}{}. This is evident from the high cross-validation scores when models are evaluated on strokes from a tabla that they have already seen during training, and the significantly poorer scores on an unseen tabla. Therefore, it is necessary to not only build a diverse dataset, but also to evaluate the instrument-independence of the proposed methods.\\

Regarding the specific task of transcribing tabla accompaniment, there has been a recent attempt at identifying strokes in the theka of a few talas \cite{sarkar2018tabla}{}. The dataset used is a diverse mix of several tabla sets, players, talas, and tempi. However, the recordings are of the thekas in their prescribed format, and limited in terms of fillers and extempore variations that we would find in a real concert. Further, while the results reported do reflect the challenges in transcribing stroke sequences as opposed to strokes recorded in isolation, there is no instrument independent evaluation reported.\\

\begin{figure}
\begin{floatrow}
\ffigbox{%
\includegraphics[width=0.45\textwidth]{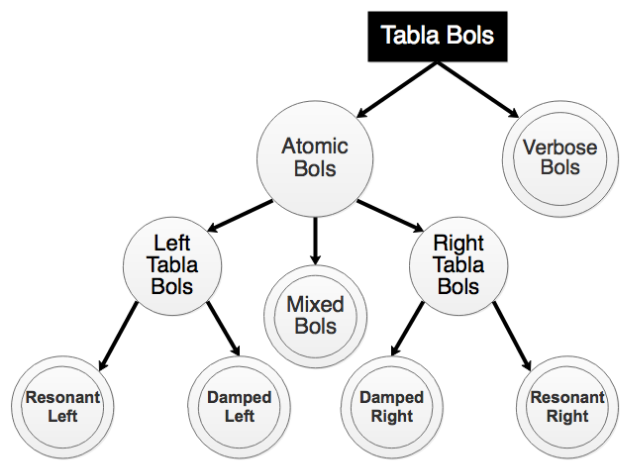}
}{%
\caption{Classification of tabla bols based on the drum that is struck and the nature of the sound produced \cite{narang2017tabla}}
\label{fig:tabla_bol_class}
}
\capbtabbox{%
\begin{tabular}{@{}ll@{}}
\toprule
\begin{tabular}[c]{@{}l@{}}Stroke \\Category\end{tabular} & \begin{tabular}[c]{@{}l@{}}Acoustic \\Characteristics\end{tabular}\\ \midrule
\begin{tabular}[c]{@{}l@{}}Damped (D) \\(\textit{Ti, Ta, Tak, Ke}, \\ \textit{Tra, Kda}) \end{tabular} & \begin{tabular}[c]{@{}l@{}} No sustained harmonics, \\only a burst of energy \\at the onset\end{tabular}\\ \\[-8pt]

\begin{tabular}[c]{@{}l@{}}Resonant \\Treble (RT) \\(\textit{Na, Tin, Tun}) \end{tabular} & \begin{tabular}[c]{@{}l@{}}One or more strong \\harmonics in the 200-2000 Hz \\range persist after the onset\end{tabular} \\ \\[-8pt]

\begin{tabular}[c]{@{}l@{}}Resonant \\Bass (RB) \\(\textit{Ghe, Dhe, Dhi}) \end{tabular} & \begin{tabular}[c]{@{}l@{}}Usually a single strong \\harmonic close to 100 Hz \\ persists after the onset\end{tabular} \\ \\[-8pt]

\begin{tabular}[c]{@{}l@{}}Resonant \\Both (B)\\(\textit{Dha, Dhin}) \end{tabular} & \begin{tabular}[c]{@{}l@{}}Both the above characteristics \\of resonant treble and bass\end{tabular} \\ \bottomrule

\end{tabular}
}{%
\caption{The four target tabla stroke categories of the transcription and their acoustic characteristics.}
\label{tab:stroke_categories}
}
\end{floatrow}
\end{figure}

\section{Problem Formulation}\label{sec:prob}
The main goal of the current work is to aid the analysis of tabla accompaniment in vocal concerts by providing musicologically meaningful stroke labels. While the outputs of a tabla transcription system can encompass all the distinct tabla bols, an important level in the taxonomy of bols is the classification into resonant and damped categories \cite{narang2017tabla} (Figure \ref{fig:tabla_bol_class}). This is based on the acoustic characteristics of the sound produced, which in turn depends on the manner in which the drums are struck - resonant sounds, produced using impulsive strikes, are harmonic and sustained, whereas damped strokes are transient and are produced by preventing the membrane from continuing to vibrate after the strike. More importantly, these categories also have musically meaningful roles in the context of accompaniment, where they help mark different sub-divisions and accent positions in the theka \cite{ajay2017rhythm}. For example, in the theka of \textit{tintal}, a 16-beat cycle, there are four sub-divisions, each of 4 beats, which are primarily distinguished based on the presence or absence of the bass resonant sound. In contrast, the different sub-divisions of the theka of \textit{ektal}, a 12-beat cycle, are distinguished based on whether they contain damped or resonant strokes.\\

\begin{figure}[!ht]
     \centering
     \begin{subfigure}[b]{0.35\textwidth}
         \includegraphics[width=\textwidth]{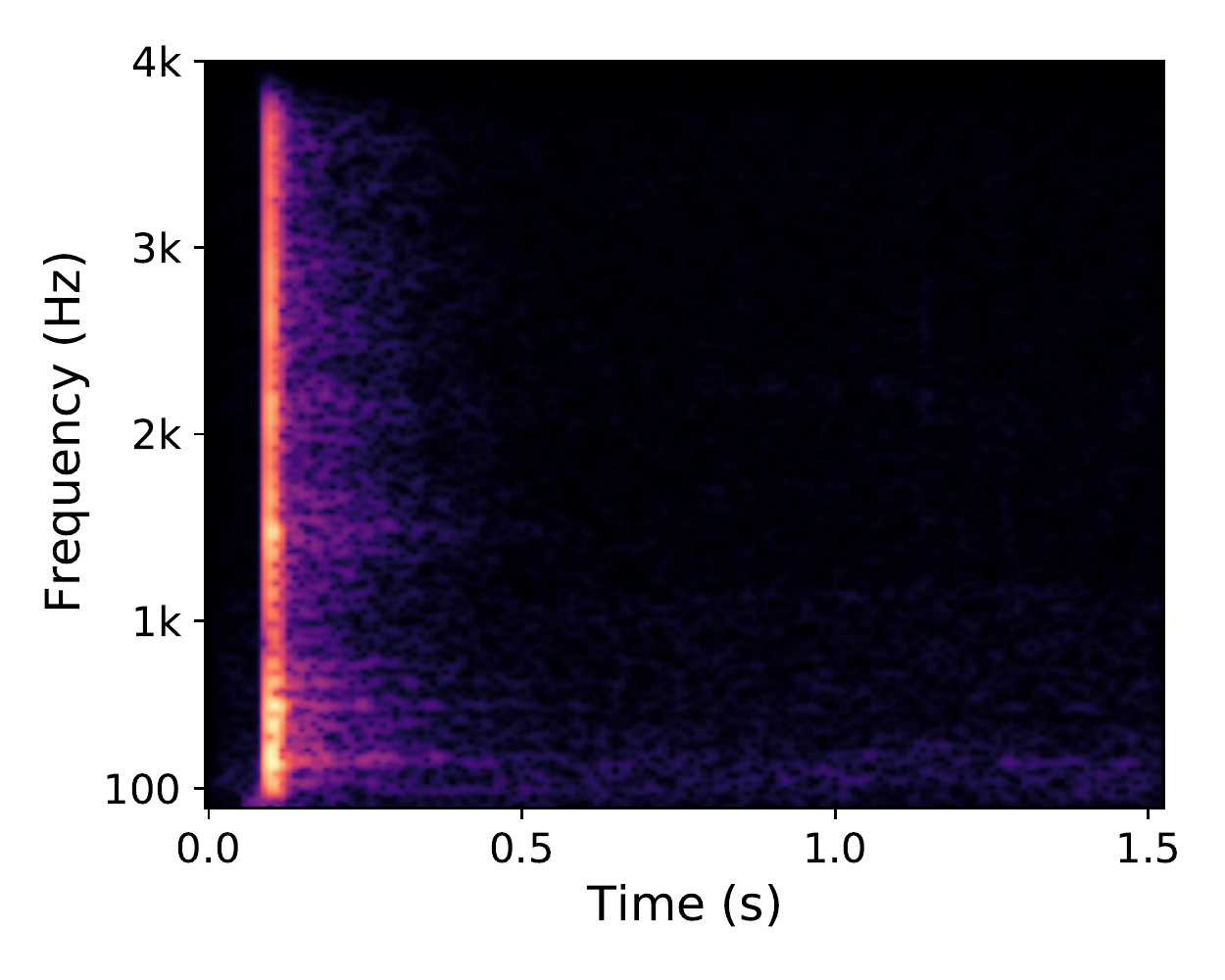}
         \caption{Damped}
         \label{fig:damped}
     \end{subfigure}
     ~
     \begin{subfigure}[b]{0.35\textwidth}
         \includegraphics[width=\textwidth]{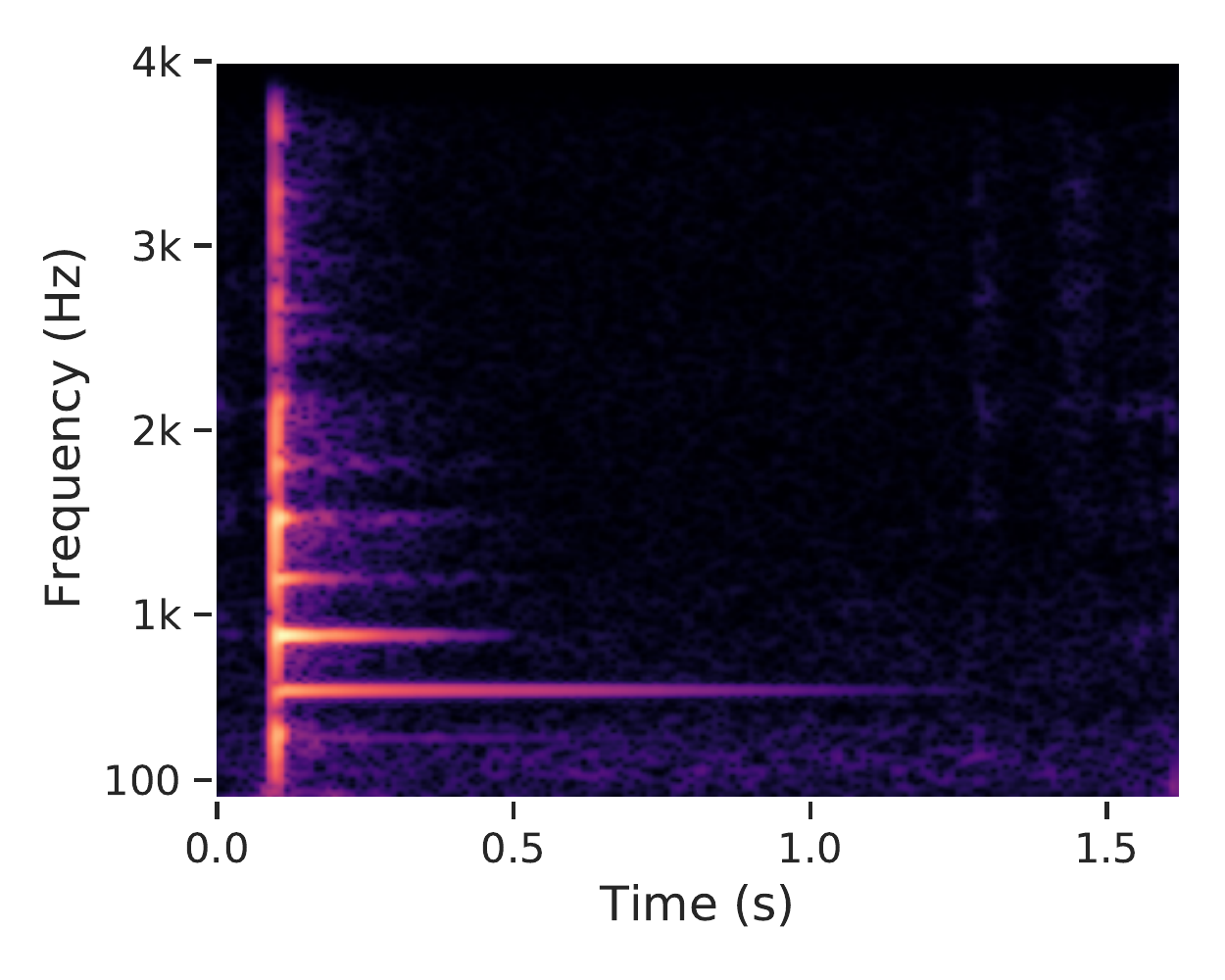}
         \caption{Resonant Treble}
         \label{fig:treble}
     \end{subfigure}
     \begin{subfigure}[b]{0.35\textwidth}
         \includegraphics[width=\textwidth]{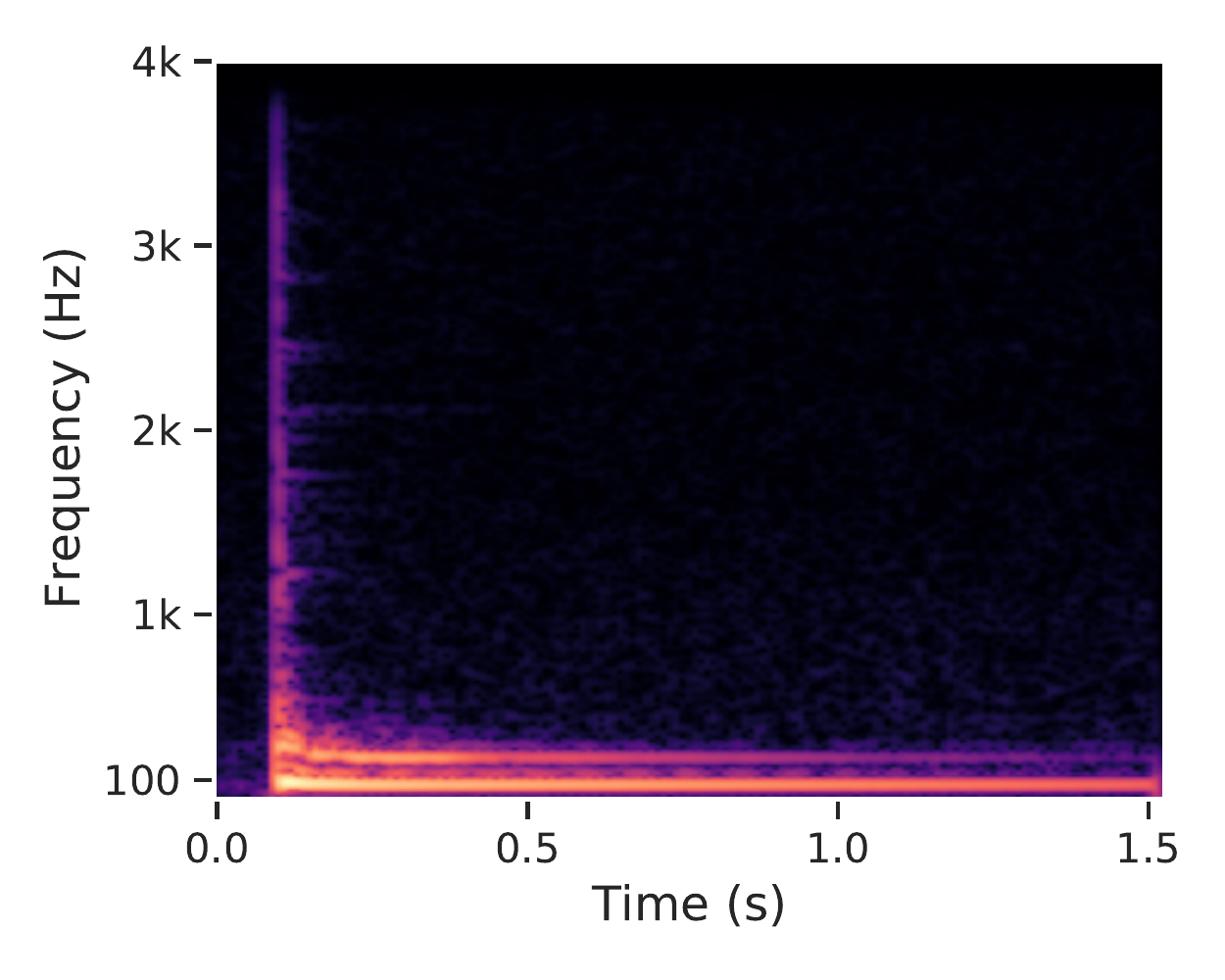}
         \caption{Resonant Bass}
         \label{fig:bass}
     \end{subfigure}
     ~
     \begin{subfigure}[b]{0.35\textwidth}
         \includegraphics[width=\textwidth]{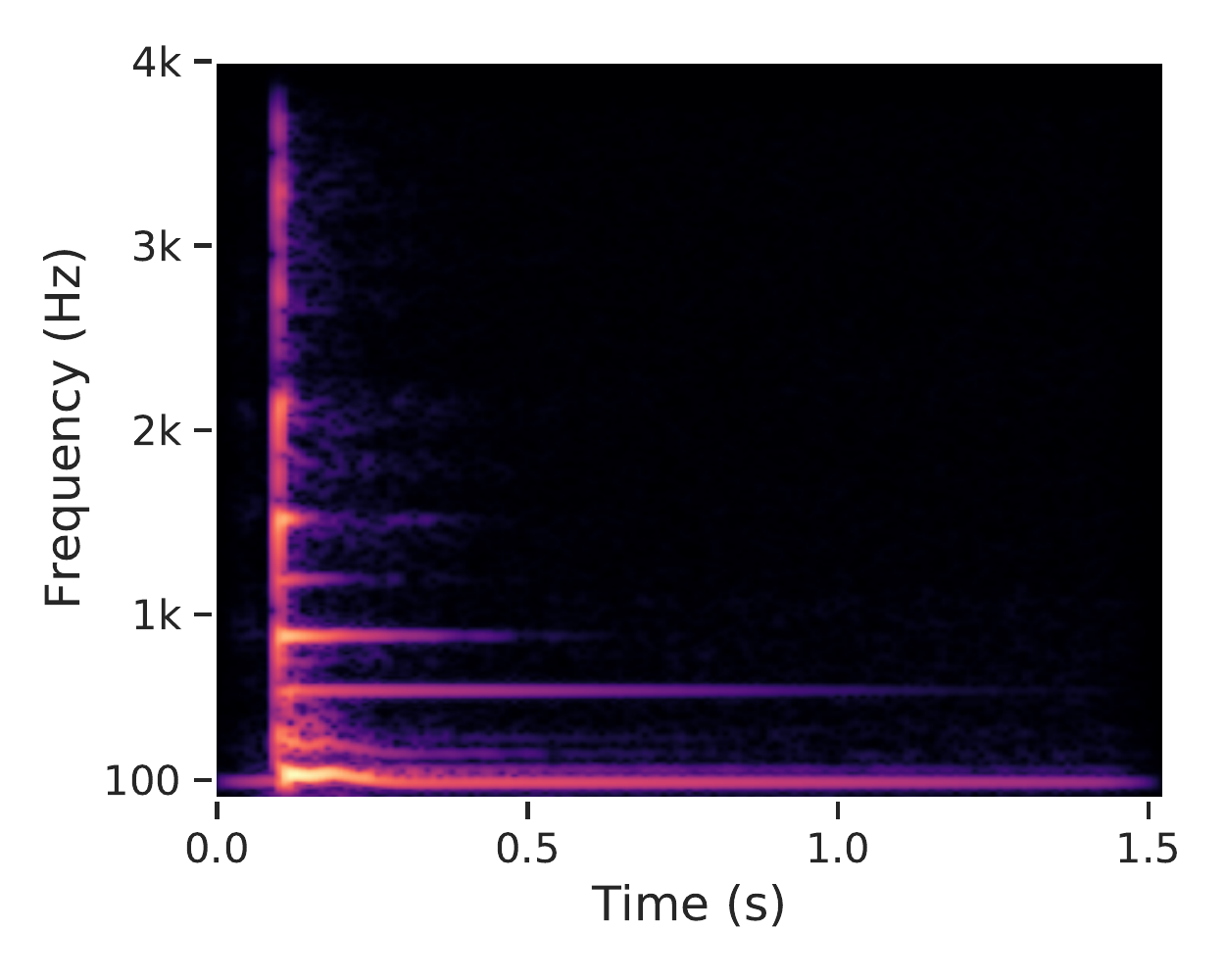}
         \caption{Resonant Both}
         \label{fig:both}
     \end{subfigure}     
        \caption{Magnitude spectrograms of a stroke from each of the four target stroke categories (spectrum shown up to 4 kHz for better visualisation of the strong harmonics whose frequency support distinguishes the resonant bass and treble strokes).}
        \label{fig:stroke_categories}
\end{figure}

We therefore look at transcribing tabla into the following four categories - \textbf{Damped}, \textbf{Resonant Treble}, \textbf{Resonant Bass}, and \textbf{Resonant Both}, based on which drum is struck and whether the sound is resonant or damped. The damped category refers to strokes that do not have a sustained decay and includes all such strokes produced either by striking one of the drums or both simultaneously (each producing a damped stroke). The resonant treble and resonant bass categories correspond to strokes that produce sustained sounds on the corresponding drums (while the other drum either remains silent or produces a damped sound). Resonant both refers to strokes that are played simultaneously on both the drums, each of which results in a sustained sound. The acoustic characteristics of each category are described in Table \ref{tab:stroke_categories}, and Figure \ref{fig:stroke_categories} shows the spectrograms for an example stroke of each category. Although a transcription into the exhaustive list of bols could be followed by then mapping the labels to any reduced set of categories, a more reasonable and also useful formulation of the problem, given the scarcity of data, is to classify each stroke directly into the required categories. \\

\section{Dataset Description}\label{sec:data_desc}
The test data for our experiments was carefully chosen to closely resemble a realistic accompaniment scenario. Since most public concert audios are not available in a multi-track format with instruments recorded in perfect isolation, it was necessary for us to build such a dataset. For this, we first obtained solo singing audios from expert singers, and then got expert tabla players to play the corresponding tabla accompaniment while listening to the pre-recorded vocals over headphones. There are a total of 10 such aligned vocal-tabla audio pairs, spanning a net duration of nearly 20 minutes, and yielding about 4500 strokes. 6 of the compositions are in \textit{madhya lay tintal} ($130-160$ BPM), and the rest are in one each of \textit{jhaptal} and \textit{bhajani theka} in \textit{madhya lay}, and \textit{tintal} and \textit{dadra} in \textit{drut lay} ($\approx 250$ BPM). The accompaniment was played by 2 different musicians for half of the tracks each, with one of the players using 2 different tabla sets based on the tonic of the singing (there are thus a total of 3 distinct tabla sets each of different tuning). All the tabla audios were recorded at a sampling rate of 44.1 kHz in quiet environments using professional microphones.\\

On the other hand, given the expensive nature of collecting such data, we obtained the data for training our classification model from three available sources of tabla solo playing, each recorded in the absence of a \textit{lehra} (melodic accompaniment), as described in Table \ref{tab:data_desc}. The \textit{PP} and \textit{AS} subsets consist of about 20 recordings each of short tabla solo compositions (5 of which are common), with each track between 10 and 30 seconds long. The \textit{AHK} subset has seven short excerpts from a solo concert recording, each between 30 and 60 seconds long. The recording, taken from YouTube\footnote{\url{https://www.youtube.com/watch?v=mEFr1Tp801M}}, is of poorer quality and at a lower sampling rate of 16 kHz. The excerpts were chosen by excluding portions of extremely fast playing that are difficult to annotate. Each of these subsets corresponds to a single unique tabla set and player.\\

\begin{table}[!t]
\centering
\begin{tabular}{@{}lllll@{}}
\toprule
Name &
  Source &
  \begin{tabular}[c]{@{}l@{}}Nature of playing\end{tabular} &
  Duration &
  \# strokes \\ \midrule
PP &
  \begin{tabular}[c]{@{}l@{}} Rohit and Rao (2018) \cite{rohit2018tabla}\end{tabular} &
  \begin{tabular}[c]{@{}l@{}} Solo compositions\end{tabular} &
  7 mins. &
  2848 \\ \\[-8pt]
AS &
  \begin{tabular}[c]{@{}l@{}}Rohit and Rao (2019) \cite{rohit2019seminar}\end{tabular} &
  \begin{tabular}[c]{@{}l@{}}Solo compositions\end{tabular} &
  5 mins. &
  1432 \\ \\[-8pt]
AHK &
  \begin{tabular}[c]{@{}l@{}}Tabla solo concert \\ audio from YouTube \end{tabular} &
  \begin{tabular}[c]{@{}l@{}}Solo compositions \\ interspersed with theka\end{tabular} &
  6 mins. &
  2400 \\ \\[-8pt] 
Test &
 \begin{tabular}[c]{@{}l@{}}Recordings made \\ for this study \end{tabular} &
 \begin{tabular}[c]{@{}l@{}} Isolated accompaniment \\ to pre-recorded vocals \end{tabular} &
 20 mins. &
 4470 \\ \bottomrule
\end{tabular}
\caption{Description of the training (\textit{PP}, \textit{AS}, \textit{AHK}) and testing datasets}
\label{tab:data_desc}
\end{table}

The \textit{PP} and \textit{AS} subsets are available along with time-aligned bol annotations. These bols were mapped according to their acoustic characteristics to the four stroke categories, for the purpose of this study. As for the \textit{AHK} and test sets, the annotation was performed manually with the help of Audacity\footnote{Audacity® software is copyright © 1999-2020 Audacity Team. The name Audacity® is a registered trademark of Dominic Mazzoni.} by the first author (a tabla student) and an intern in the same lab. To aid the manual annotation process, stroke onsets were first determined automatically using an automatic onset detection algorithm (Section~\ref{sec:onset_det}). In order to make up for missed soft onsets, a fairly low threshold was used while picking the peaks in the onset detection signal. This resulted in some false alarms, but ensured that soft onsets were also detected, thus requiring the annotator to not add any onsets, but only remove the false alarms and appropriately label the correct onsets.\\

\begin{figure}
    \centering
    \includegraphics[width=0.6\textwidth]{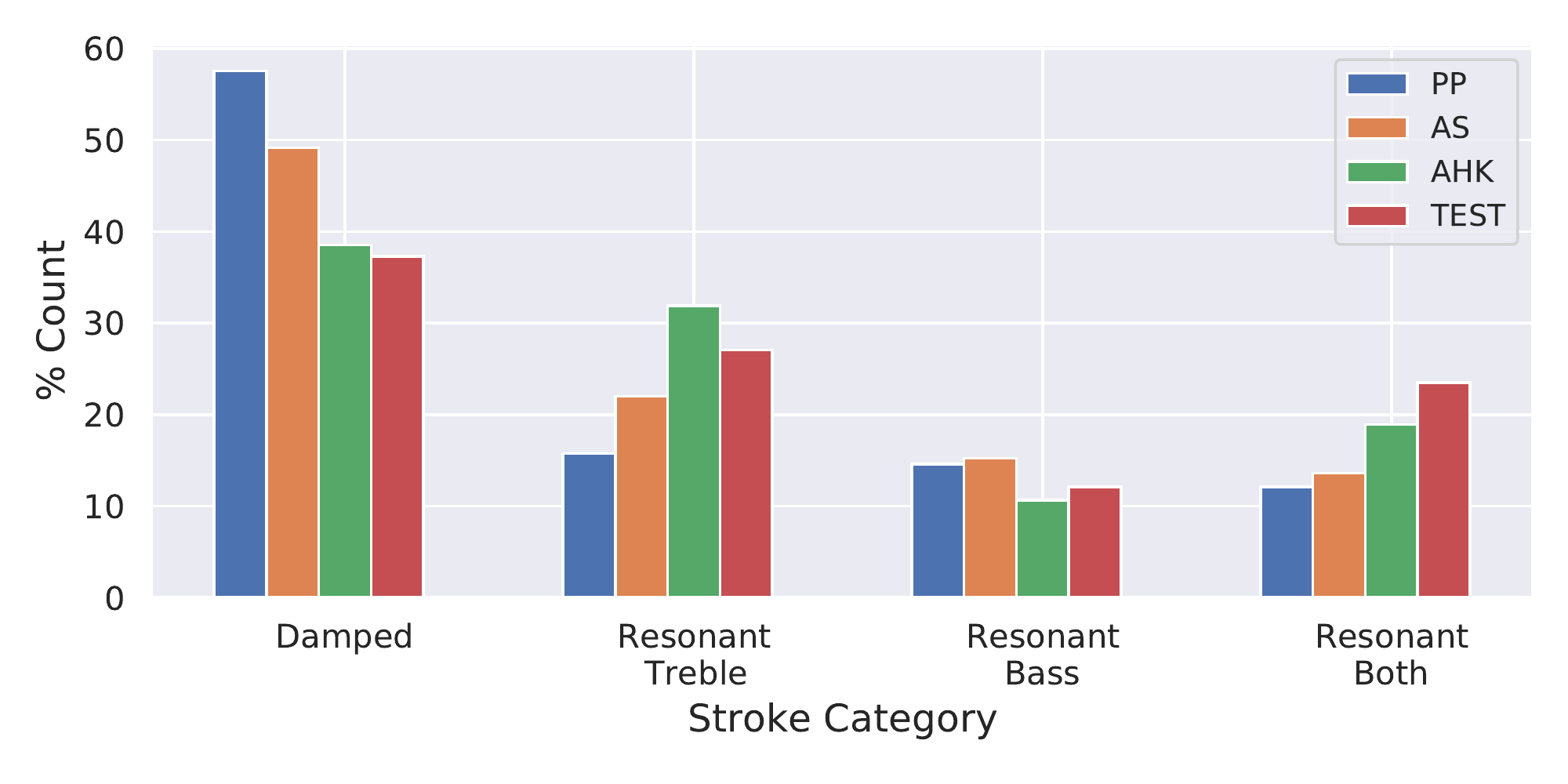}
    \caption{Distribution (\% count) of strokes across categories in the training and testing sets.}
    \label{fig:data_desc}
\end{figure}

The distribution of strokes across the four categories in each of these subsets is shown in Figure~\ref{fig:data_desc}. Comparing these with the test set, we see a few similarities as well as differences. In all the subsets (train and test), the damped category is the most populated and resonant bass is the least populated. The high count of damped strokes can be attributed, in the case of solo playing, to the nature of solo compositions - they contain more damped strokes to allow playing at high speeds, and in the case of accompaniment, to fillers and expressive embellishments. Further, in the test set, the resonant both category has a higher count than any of the training subsets, possibly because of the dominant nature of such strokes in accompaniment playing (for instance, the theka for \textit{tintal} has 12 out of its 16 strokes of this kind).\\

\section{Methods}
In the present study, we follow the same `segment and classify' approach, where stroke onsets are first determined and the labels are then assigned to the segmented strokes. Each segment is the duration from the onset of a stroke to the onset of the subsequent stroke. We extract features on these segments and provide the feature vectors to a random forest classifier model to predict the stroke category. All the tabla audios (training and testing) are first down-sampled to 16 kHz, since the \textit{AHK} set is not available at 44.1 kHz.

\subsection{Onset Detection}\label{sec:onset_det}
Given the well-studied nature of onset detection, especially for percussive sounds, we make use of a common onset detection method called the High Frequency Content, and experiment with the post-processing hyperparameters to arrive at the best settings for our data. This method involves calculating the frequency-weighted sum of the magnitude spectrum to indicate the onset strength in each frame \cite{bello2005onset}. The implementation is as provided in the software library \textit{Essentia} \cite{essentia2013}, with the window and hop size values set to 25 ms and 5 ms, respectively. The post-processing step involves smoothing the onset detection signal and then picking the peaks above a certain threshold \cite{brossier2004onset}. The size of the moving average filter (\textit{delay}) used for smoothing and the threshold value for peak-picking (\textit{alpha}) are hyperparameters that we experiment with. We first iteratively evaluate each of the methods on the training set using a grid of values for \textit{alpha} and \textit{delay}, and pick the pair of values that results in the highest f-score. Then, we keep these values fixed and evaluate the onset detector on the test set.\\

\subsection{Feature Extraction}
The features we use in this work are modeled on those used in previous tabla transcription methods (as reviewed in Section \ref{sec:background}), along with a few important modifications and additions. The total set of features, 49 in all, is given in Table~\ref{tab:feat_set}. Some of the features are calculated separately on band-passed versions of the signal. Given that our four-way target stroke classes are based closely on the differences in the resonant characteristics of each drum, we decide on the following ranges for the bands: 50 - 200 Hz to capture the base and 200 - 2000 Hz to capture the treble harmonics. These ranges are kept broad in order to capture a wide F0 range of each drum, similar to the bands used in \cite{gillet2003tabla}{}. They could, however, be tuned based on the tabla as well, by first identifying the range of harmonic frequencies of each drum, as done previously (for the treble drum alone) in \cite{narang2017tabla}{}. This could be achieved by analysing a few resonant strokes of each drum, which, in a testing scenario, could be provided manually.\\

\begin{table}[!h]
\centering
\begin{tabular}{@{}llll@{}}
\toprule
Category                  & Feature                       & \multicolumn{2}{l}{Count} \\ \midrule
\multirow{4}{*}{Spectral} & 
$\left.\begin{tabular}{@{}l@{}}
      Spectral Centroid \\
      Skewness \\
      Kurtosis \\
      \phantom{Aardvark}\\[-\normalbaselineskip]
    \end{tabular}\hspace{1cm}\right\}$ &
    2 (mean, stdev.) \\ \\[-4pt]
                          & MFCC                          & 13 (mean)   \\[+7pt]
                          & \begin{tabular}[l]{@{}l@{}}Flux (onset strength)\\ Energy\end{tabular} & 
$\left.\begin{tabular}{@{}l@{}}
      1 (max.) \\
      3 (sum, mean, stdev.) \\
      \phantom{Aardvark}\\[-\normalbaselineskip]
    \end{tabular}\right\}$ 
    x 2 (bass, treble) & \\ \\[-4pt]
                          
\midrule
\multirow{8}{*}{Temporal} & Log Attack Time & 1 \\
                          & Temporal Centroid & 1 \\
                          & Zero Crossing Rate & 2 (mean, stdev.) \\  \\[-4pt]                          & \begin{tabular}[l]{@{}l@{}}  Early Decay Rate \& Intercept \\ Late Decay Rate \& Intercept \\ $R^2$ \\ Spline Knot Location \end{tabular} & 
$\left.\begin{tabular}{@{}l@{}}
      2 \\
      2 \\
      1 \\
      1 \\
      \phantom{Aardvark}\\[-\normalbaselineskip]
    \end{tabular}\right\}$
    x 2 (bass, treble) & \\ \\[-4pt]
\midrule 
Delta    & \begin{tabular}[l]{@{}l@{}}Sum and Mean Energy \\ Late Decay Rate \end{tabular} & 
$\left.\begin{tabular}{@{}l@{}}
      2 \\
      1 \\
      \phantom{Aardvark}\\[-\normalbaselineskip]
    \end{tabular}\right\}$
    x 2 (bass, treble) & \\ \\[-4pt]
\bottomrule \\[-4pt]
& \begin{tabular}[r]{@{}r@{}}\textbf{Total}\end{tabular} & \textbf{49} \\ \\[-4pt]
\end{tabular}
\caption{The set of features used for stroke classification.}
\label{tab:feat_set}
\end{table}

All the spectral features, and the temporal zero-crossing rate feature, are calculated frame-wise on overlapping short-time frames of size 25 ms, with a hop of 5 ms, over the entire stroke segment. The shape-related spectral features (spectral centroid, skewness, and kurtosis) are summarised using the mean and standard deviation, MFCC using the mean, flux using the maximum, and energy using the sum, mean, and standard deviation. The other two temporal features - log attack time and temporal centroid are calculated from the amplitude envelope of the entire time-domain audio signal of the stroke segment. The implementation provided by \textit{Essentia} is used for each of these features.\\

\begin{figure}[!t]
     \centering
     \begin{subfigure}[b]{0.4\textwidth}
         \centering
         \includegraphics[width=\textwidth]{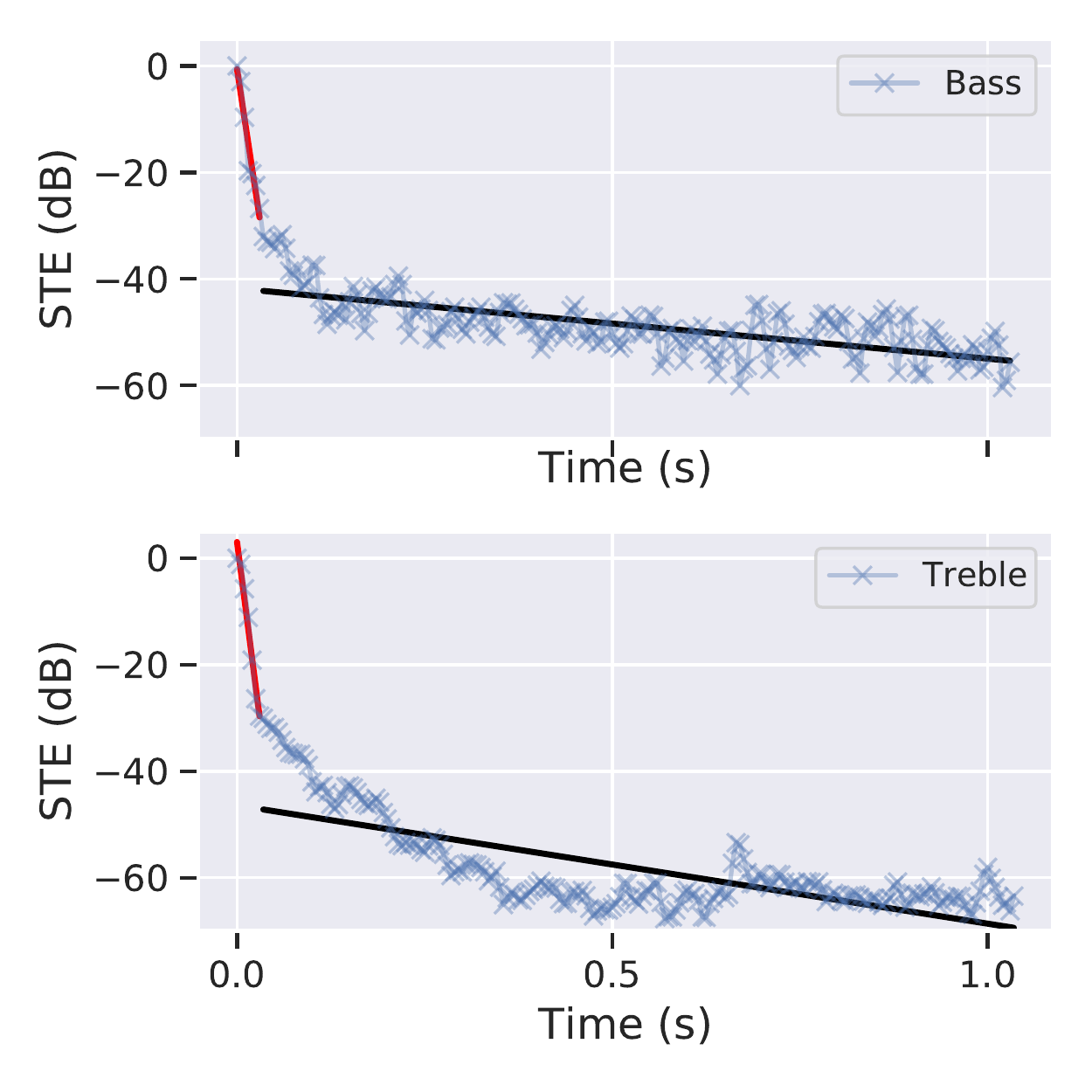}
         \caption{Damped}
         \label{fig:damped_ste}
     \end{subfigure}
     ~
     \begin{subfigure}[b]{0.4\textwidth}
         \centering
         \includegraphics[width=\textwidth]{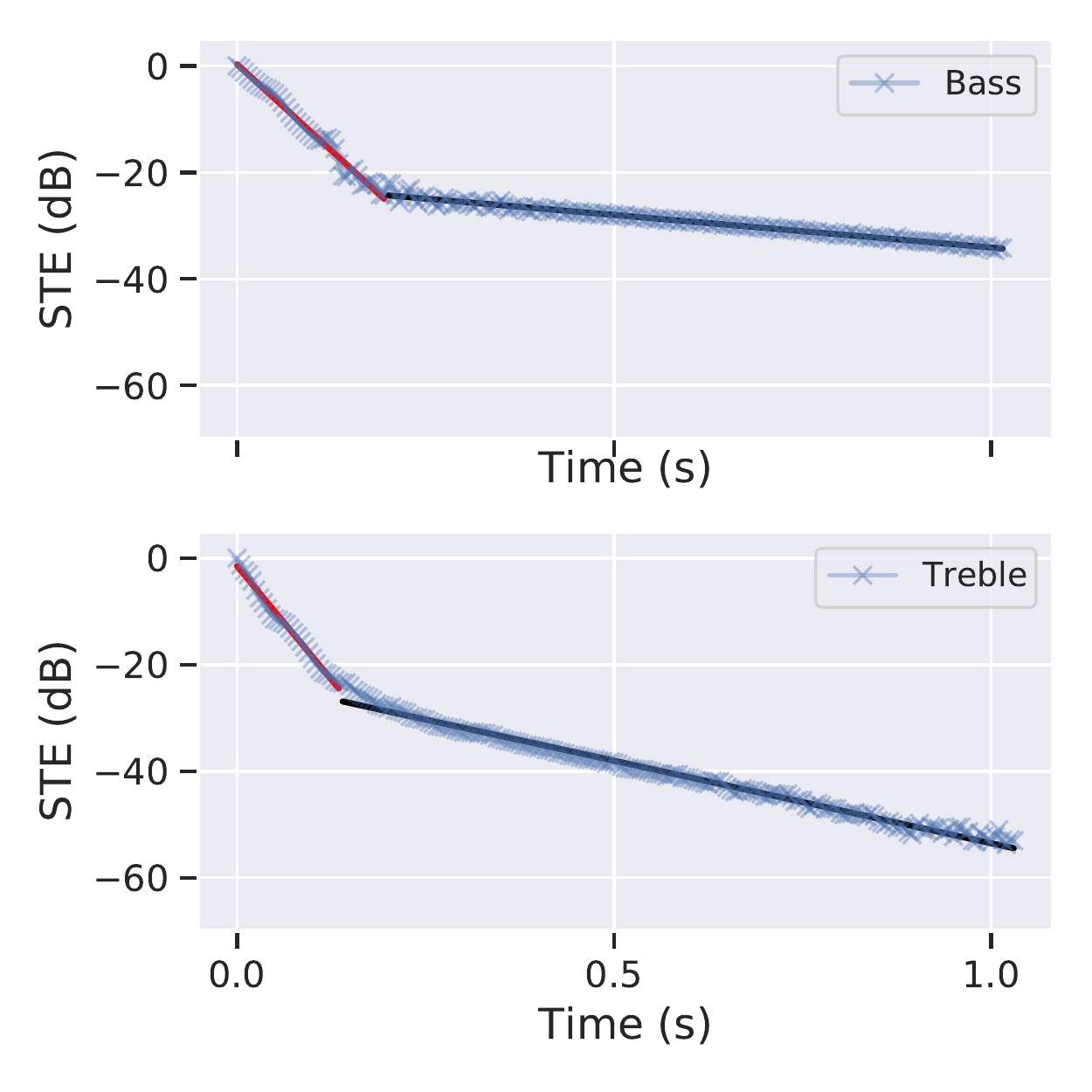}
         \caption{Resonant Both}
         \label{fig:both_ste}
     \end{subfigure}
     
        \caption{Band-wise log-transformed short-time energy envelopes (marked by `x') of a stroke from (a) Damped and (b) Resonant Both categories, along with the best spline fit (solid lines).}
        \label{fig:decay_rates}
\end{figure}

For the remaining temporal features, the short-time energy envelope serves as the base. A logarithmic transformation is first applied to make the decay portion (the part beyond the peak) of the envelope more linear. By observing a few plots of band-wise log-scaled short-time energy envelopes (Figure \ref{fig:decay_rates}), we see that the decay portion usually consists of two parts (early and late), and the differences between the stroke categories seem to be based on the rates of these decays. For instance, the decay patterns in both the bands are quite similar for the damped stroke, with the late portion mostly corresponding to noise, whereas in the resonant both stroke, the bands contain smooth decays at different rates. Therefore, to capture the distinctive nature of the decay, a linear spline model is fit to the decay portion.
The `knot' location for the fit, which is the point of inflexion, is automatically estimated as the point that yields the combined best fit for the two linear segments, in the sense of the harmonic mean of the $R^2$ values of each fit. The slopes of both the line segments, their y-intercepts, the harmonic mean of the $R^2$ values, and the knot location are all used as features. The terms `early' and `late' are used to refer to the two line segments in terms of their relative positions. The use of decay-related parameters as features is reported in \cite{narang2017tabla}{}. However, a key difference here is the use of a linear spline fit to the log-scaled envelope instead of a single exponential fit to the un-scaled envelope as done previously. \\

Finally, to help determine whether the measured characteristics of a stroke are not corrupted by possible overlap from a previous stroke, we incorporate delta features.
These are calculated as the difference between the values in a given and its previous stroke for the following six features - bass and treble late decay rates, and the sum and mean of base and treble short-time energy values.\\

\subsection{Stroke Classification}
For stroke classification, we use a random forest classifier, with the implementation provided in the scikit-learn library \cite{sklearn}{}. A random forest classifier is an ensemble of decision trees in which the output is determined by a majority vote across all the trees. The main motivation behind its design is to prevent the overfitting that is commonly observed in the case of a single decision tree. This model is also useful because it lends itself well to an analysis of the important features in the input data (given by the `feature\_importances\_' attribute of a trained model). The input to the model is a feature vector comprising all the features listed in Table \ref{tab:feat_set}, extracted on a single stroke segment, and the target is one of the four stroke categories. For our initial experiments, we set the number of trees in the forest to 200, the `max\_features' parameter to `sqrt', and the `bootstrap' parameter to `True' (Section \ref{sec:feat_sel_hyp_tun}).

\subsubsection{Feature Selection and Hyperparameter Tuning}\label{sec:feat_sel_hyp_tun}
The subset of features that are most relevant and non-redundant can be obtained via a recursive feature elimination (RFE) strategy. This method uses a classifier like the random forest to first estimate the relative feature importances, and then removes the least important ones recursively until a desired number of features is obtained. In order to obtain the optimum number of features, it is performed in a cross-validation setting on the training set, where a second classifier model is trained and evaluated on separate folds to obtain the average cross-validation accuracy at a given number of selected features. By repeating this for different values of the number of selected features and recording the average classification score in each case, a suitable point on this curve can be chosen while trading off performance for complexity.\\

An important step in extracting the best performance out of a classifier is the optimization of its hyperparameters. In the case of the random forest, these determine various aspects of the model complexity and help control the trade-off between bias and variance - a more complex model increases the variance while a simpler model increases the bias. In the present work, we look at optimizing the following hyperparameters - number of trees, maximum allowed depth of each tree, size of the subset of features given to each tree (also called `max\_features'), and the use of bootstrapping in sampling subsets of the training data. The optimization is carried out by sampling 40 random combinations of these hyperparameters from a grid of values and evaluating each combination in terms of the average cross-validation score on the training set. The use of cross-validation here is intended to not bias the model towards our test set of accompaniment audios in the process of tuning the model hyperparameters.

\subsubsection{Data Augmentation and Balancing}
A recurring problem that has been noted by authors in some of the previous work on tabla transcription is the scarcity and limited diversity of available datasets. The process of building large, accurately annotated datasets is, in general, extremely cumbersome, but especially so in the case of instruments like tabla, which cannot be synthesized realistically using existing software or electronic instruments (most software only has recorded samples of short loops). This motivates the use of effective data augmentation strategies where available labeled data is transformed in different ways that do not change the validity of the labels, thus offsetting to an extent the need to collect more data.\\

For our present work, we investigate the following three general data balancing and augmentation methods - repeating the data samples (repeated oversampling), interpolating in the feature space, and pitch-shifting the audios. For pitch-shifting, we use the function provided in the software library \textit{librosa} \cite{librosa}{}, and only consider the three resonant stroke categories (since damped strokes do not contain a pitched component). Each stroke's audio signal is pitch-shifted by each of 4 semitone levels: \{-0.5, -0.25, 0.25, 0.5\}. This rather small range of shifts compares well with the allowed range for a high-pitched tabla of a small-diameter, which is what the tabla sets in our training set resemble \cite{courtney2016tabla}.\\

Given that our dataset is also highly imbalanced in terms of the distribution of examples across the four categories, we use the methods of repeated oversampling and interpolation as required to achieve a balanced distribution. That is, we augment samples in all the categories except the majority class, until a balance is obtained. For the interpolation, we use an algorithm called \textit{SMOTE} (Synthetic Minority Oversampling Technique) \cite{chawla2002smote}{}. This uses a k-nearest neighbours method to first estimate the \textit{k} samples nearest to a given data sample, after which, between every pair of the given sample and a neighbour, a new feature vector is sampled by linearly interpolating between the feature vector pair. The implementation for this is as provided in the library \textit{imbalanced-learn} \cite{imblearn}{}.\\

\section{Results and Discussion}\label{sec:results}
For the stroke onset detection task, the commonly used metrics for evaluation are the precision, recall, and f-score, which are calculated by determining the correct and incorrect onset predictions. A correctly detected onset (`hit') is one that lies within a tolerance window ($50 ms$ wide) about an unmatched ground truth onset \cite{raffel2014mireval}. The average f-score is calculated as the mean value across all the tracks in a dataset. After the hyperparameter tuning, we obtain a best f-score of 0.972 on the training set, while on the test set, we obtain a comparably high f-score of 0.965.\\

To report stroke classification performance, we use the accuracy as well as  average f-score, calculated by averaging the per-class f-score value. We perform cross-validation in a `Leave-One-Tabla-Out' (LOTO) format, consisting of 3 folds, one each corresponding to each of the 3 tabla sets in the training dataset. This prevents the samples from a given tabla from being present in both the training and evaluation folds and therefore results in a more realistic evaluation of performance. The LOTO-CV results are reported for each of the 3 tabla sets in Table \ref{tab:stroke_class_perf} which also provides the test set performance from a model trained on all the 3 solo tabla training sets. We present different data augmentation and balancing strategies, including some combinations of the two. The baseline method refers to training the classifier on just the originally available dataset.
Among the different methods, the use of pitch-shifting leads to a significant rise in the test set score, while resulting in no improvement over the baseline in the average cross-validation f-score. On the other hand, the other two data balancing strategies (repeated oversampling and SMOTE) improve the average CV scores, mainly due to an improvement on the \textit{AS} set alone, while not positively affecting the test set scores (even when performed along with pitch-shifting). This indicates that the latter two methods are perhaps resulting in an overfit on the training data, whereas the pitch-shifting method is helping the model generalise better. Further, among the three training sets, we see a consistently poorer CV performance on the \textit{AHK} set, while the other two are comparable.\\

\begin{table}[!t]
\centering
\begin{tabular}{@{}ccccccccccc@{}}
\toprule
\multirow{3}{*}{Method} &
  \multicolumn{8}{c}{Cross-validation scores} &
  \multicolumn{2}{c}{\multirow{2}{*}{\textbf{\begin{tabular}[c]{@{}c@{}}Test set \\ scores\end{tabular}}}} \\ \cmidrule(lr){2-9}
 &
  \multicolumn{4}{c}{Accuracy} &
  \multicolumn{4}{c}{F-score} &
  \multicolumn{2}{c}{} \\ \cmidrule(l){2-11} 
 &
  \textit{PP} &
  \textit{AS} &
  \textit{AHK} &
  Mean &
  \textit{PP} &
  \textit{AS} &
  \textit{AHK} &
  Mean &
  Acc. &
  F-sc. \\ \midrule


\begin{tabular}[c]{@{}c@{}}Baseline \\ (no augmentation)\end{tabular} &
  0.80 &  0.73 &  0.57 &  0.70 &  0.70 &  0.58 &  0.49 &  0.59 &  0.59 &  0.50 \\ \\[-8pt]
\begin{tabular}[c]{@{}c@{}}Repeated \\ oversampling\end{tabular} &
  0.80 &  0.74 &  0.57 &  0.70 &  0.70 &  0.62 &  0.49 &  0.60 &  0.59 &  0.51 \\ \\[-8pt]
SMOTE &
  0.79 &  0.76 &  0.57 &  \textbf{0.70} &  0.70 &  0.67 &  0.49 &  \textbf{0.62} &  0.59 &  0.51 \\ \\[-8pt]
Pitch-shifting &
  0.76 &  0.66 &  0.55 &  0.66 &  0.69 &  0.58 &  0.49 &  0.59 &  \textbf{0.65} &  \textbf{0.60} \\ \\[-8pt]
\begin{tabular}[c]{@{}c@{}}Pitch-shifting \\ + Rep. Oversampling\end{tabular}&
  0.76 &  0.67 &  0.56 &  0.66 & 0.69 &  0.61 &  0.49 &  0.60 &  0.64 &  0.59 \\ \\[-8pt]
\begin{tabular}[c]{@{}c@{}}Pitch-shifting \\ + SMOTE\end{tabular}&
  0.76 &  0.70 &  0.56 &  0.67 & 0.69 &  0.62 &  0.48 &  0.60 &  0.64 &  0.58 \\ \bottomrule
\end{tabular}
\caption{Leave-One-Tabla-Out cross-validation scores (on each fold and the mean) and the test set scores for stroke classification using the various augmentation methods (the highest mean CV and test scores obtained are highlighted in bold).}
\label{tab:stroke_class_perf}
\end{table}

\begin{figure}[!ht]
     \centering
     \begin{subfigure}[b]{0.45\textwidth}
         \centering
         \includegraphics[width=\textwidth]{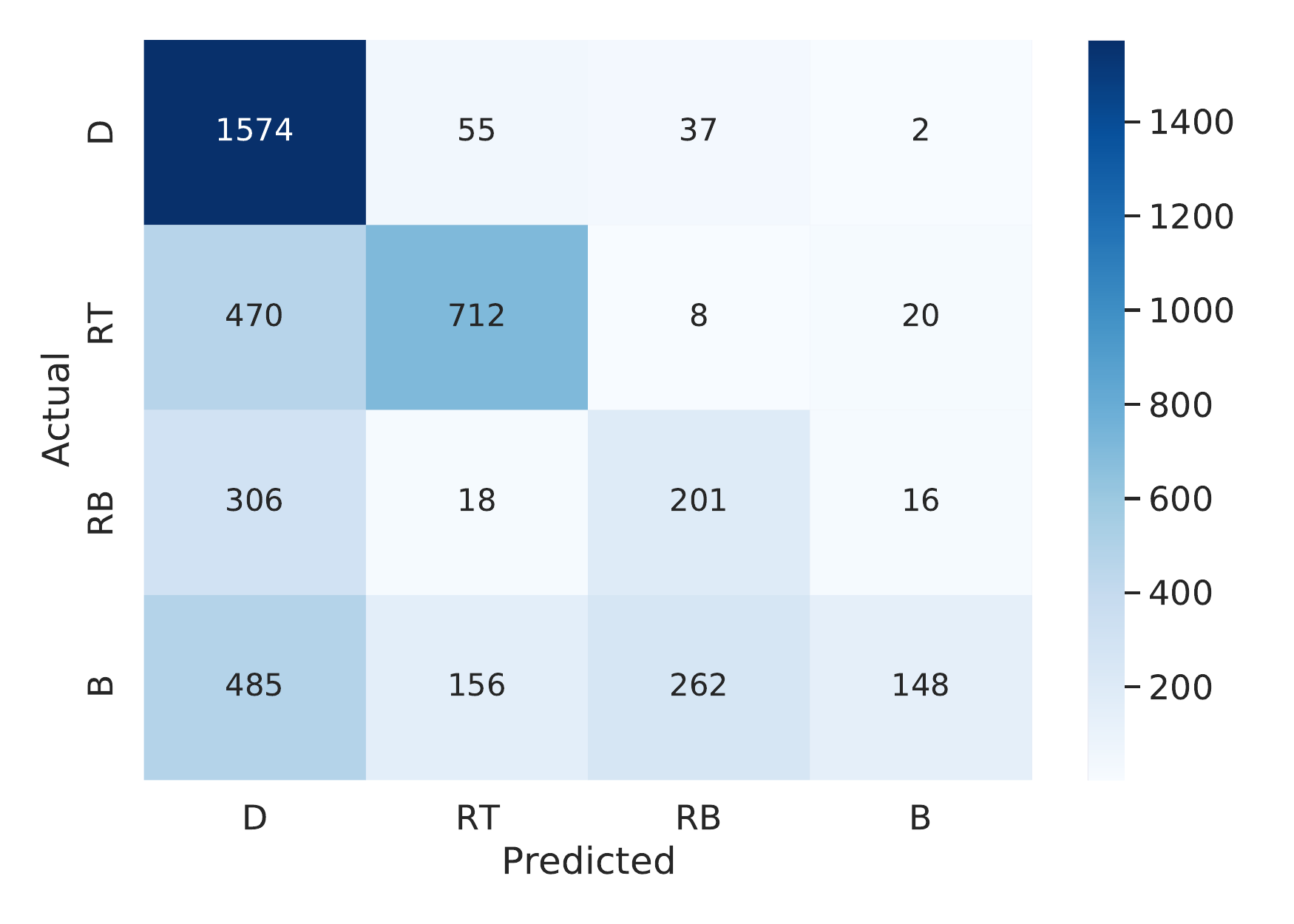}
         \caption{Baseline}
         \label{fig:cm_baseline}
     \end{subfigure}
     \hfill
     \begin{subfigure}[b]{0.45\textwidth}
         \centering
         \includegraphics[width=\textwidth]{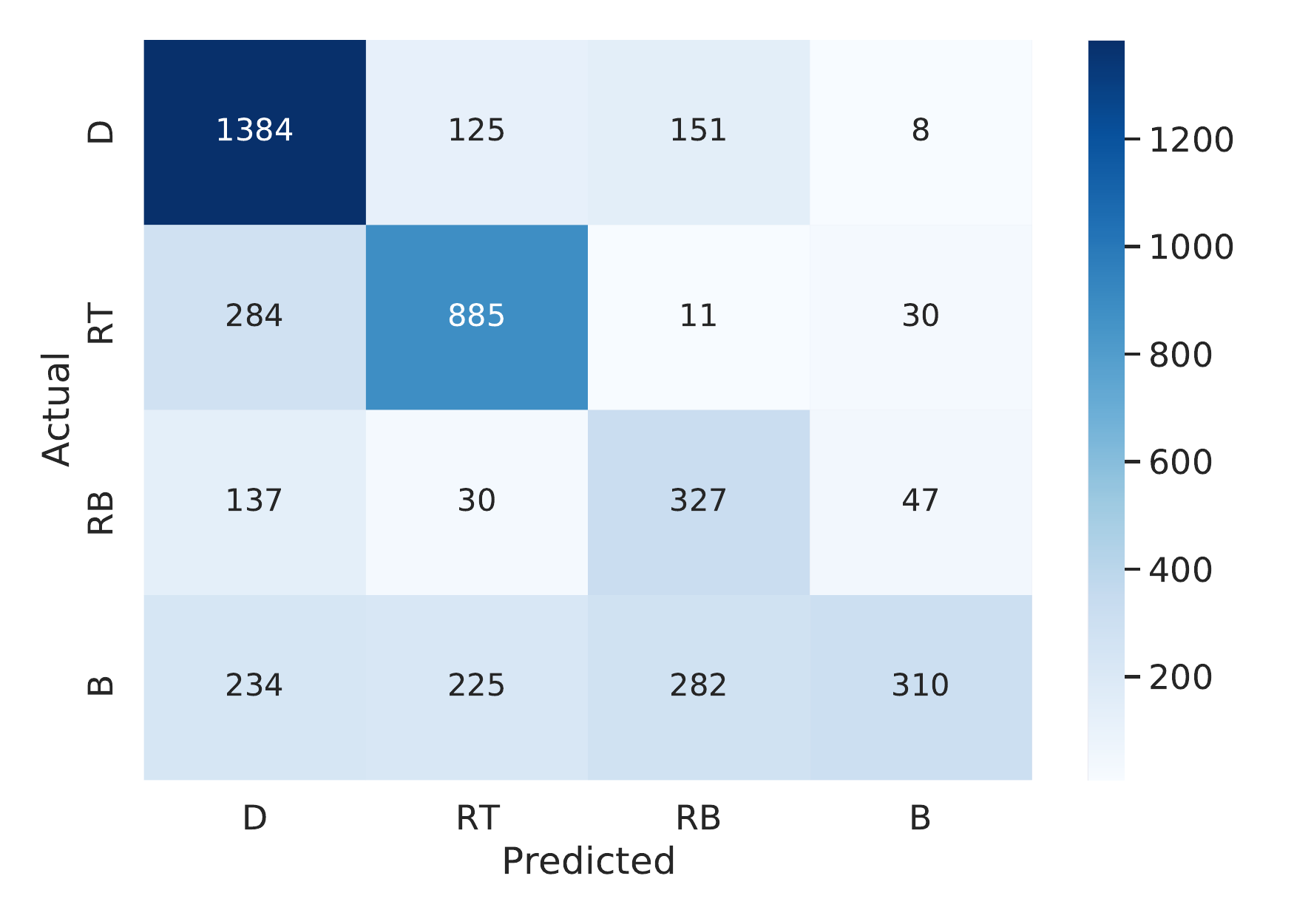}
         \caption{Pitch-shifting}
         \label{fig:cm_pitch}
     \end{subfigure}     
        \caption{Confusions in the test set stroke classification with (a) baseline method and (b) pitch-shifting augmentation. (D: `Damped', RT: `Resonant Treble', RB: `Resonant Bass', B: `Resonant Both')}
        \label{fig:conf_mats_test}
\end{figure}

From Figure \ref{fig:conf_mats_test}, which shows the confusions in stroke label predictions on the test set, we see that the model is clearly biased towards the `damped' class in the baseline method due to the large class imbalance in the training set. With the use of pitch-shifting augmentation, the errors in all the resonant categories reduce, but at the cost of a slight increase in errors in the damped category. Most of the remaining errors are in the resonant bass and resonant both categories.
\begin{table}
\centering
\begin{tabular}{@{}cccccccccc@{}}
\toprule
\multirow{3}{*}{Method} &
  \multicolumn{8}{c}{Stroke category} &
  \multirow{3}{*}{\begin{tabular}[c]{@{}c@{}}Balanced \\ accuracy\end{tabular}} \\ \cmidrule(lr){2-9}
 &
  \multicolumn{2}{c}{D} &
  \multicolumn{2}{c}{RT} &
  \multicolumn{2}{c}{RB} &
  \multicolumn{2}{c}{B} &
   \\ \cmidrule(lr){2-9}
               & Prec. & Rec. & Prec. & Rec. & Prec. & Rec. & Prec. & Rec. &      \\ \midrule
Baseline       & 0.82  & 0.93 & 0.60  & 0.95 & 0.86  & 0.20 & 0.63  & 0.29 & 0.59 \\
Pitch-shifting & 0.93  & 0.69 & 0.48  & 0.96 & 0.71  & 0.36 & 0.45  & 0.39 & 0.60 \\
\begin{tabular}[c]{@{}c@{}}Pitch-shifting \\ + SMOTE\end{tabular} &
0.92 &  0.75 &  0.53 &  0.96 &  0.74 &  0.39 &  0.50 &  0.43 &  0.63 \\ \bottomrule
\end{tabular}
\caption{Class-wise validation scores on the AS set in the LOTO-CV. The balanced accuracy is the average per-class accuracy (suitable for imbalanced datasets).}
\label{tab:stroke_class_perf_as}
\end{table}
To help explain the previously noted drop in CV accuracies in the course of pitch-shifting based augmentation, we look at the class-wise precision and recall values on the \textit{AS} set alone when it is evaluated in the LOTO-CV procedure (Table \ref{tab:stroke_class_perf_as}). Also reported is the `balanced accuracy' - the average per-class accuracy, which is more suitable for evaluating the performance on imbalanced datasets. The significant drop in the recall value for the damped stroke with the use of pitch-shifting, coupled with its higher count in the dataset, brings out the reason for a drop in the overall accuracy. However, the balanced accuracy improves with the use of pitch-shifting augmentation, and further, with SMOTE. Further, the opposite trends in the precision and recall values for the resonant stroke categories with the use of pitch-shifting explain the lack of improvement in the CV f-score.\\

In Table \ref{tab:stroke_class_perf_prev}, we offer a comparison of our baseline cross-validation results (without any augmentation) with a few relevant tabla transcription results from the literature. We also evaluate our method using a random 3-fold CV (as opposed to the LOTO-CV reported in Table \ref{tab:stroke_class_perf}) to aid this comparison. The dataset size, number of tabla sets, number of target classes, and modes of evaluation are all different in each of the methods. Some of the scores are reported as ranges because the average accuracy reported by Chordia is calculated over a number of individual results on a few subsets of the data. Some of these subsets include single tabla sets out of the total of 7, contributing sometimes to the high average accuracy. Hence, the ranges reported here are an attempt to cover the whole set of accuracies reported under the different conditions. A similar approach was taken to arrive at the accuracy range reported for our method as well. Although a direct comparison is difficult due to the larger target class set in the previous methods (comprising all the distinct bols), a key observation is the huge range of LOTO-CV score reported by Chordia \cite{chordia2004tabla}{}, pointing again to the larger problem of generalisation to unseen instruments.\\

\begin{table}
\centering
\begin{tabular}{@{}ccccc@{}}
\toprule
Method &
  \begin{tabular}[c]{@{}c@{}}\# Target \\ classes\end{tabular} &
  \begin{tabular}[c]{@{}c@{}}\# strokes, \\ tabla-sets\end{tabular} &
  \begin{tabular}[c]{@{}c@{}}Random \\ CV acc.\end{tabular} &
  \begin{tabular}[c]{@{}c@{}}LOTO-CV\\ acc.\end{tabular} \\ \midrule
Ours          & 4     & 6678, 3  & 0.71 - 0.92 & 0.57 - 0.80 \\
Sarkar et. al. \cite{sarkar2018tabla} & 10 & 3964, 7 & 0.85 & -\\
Gupta et. al. \cite{gupta2015tabla} & 18    & 8200, 1  & 0.66        & -           \\
Chordia \cite{chordia2004tabla}       & 10-16 & 16384, 7 & 0.77 - 0.94 & 0.15 - 0.95 \\ \bottomrule
\end{tabular}
\caption{Comparing the cross-validation results of our method with previous tabla transcription studies.}
\label{tab:stroke_class_perf_prev}
\end{table}

\begin{figure}[!t]
    \centering
    \includegraphics[width=0.55\textwidth]{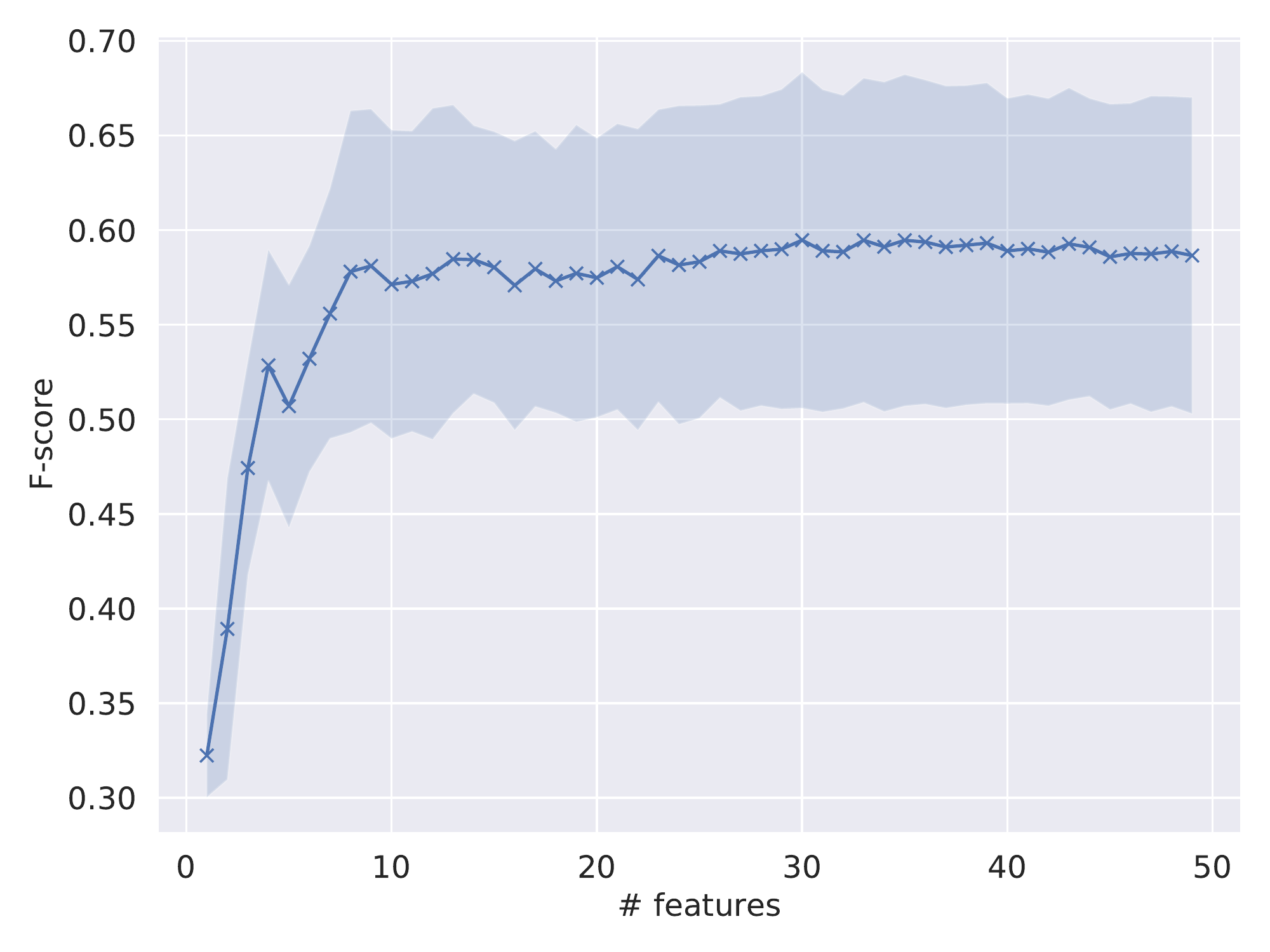}
    \caption{Plot of average LOTO-CV f-score versus the number of features selected in the recursive feature elimination procedure (RFE). The shaded region represents the standard deviation of the score across the CV folds.}
    \label{fig:rfe}
\end{figure}

Next, we look at the results from the feature selection and model hyperparameter tuning. The plot of average LOTO-CV f-score versus the number of features shows a rise till about 30 features, after which it first plateaus and then dips slightly towards the end (Figure \ref{fig:rfe}). The RFE method was then applied to the entire training set in order to pick the best set of 30 features, and the model tuning was performed only using this subset of all features. At the outset, there was no significant improvement in the average CV f-score compared to the results of Table \ref{tab:stroke_class_perf}. A decrease in the number of trees was not found to cause a significant drop in the f-score, perhaps due to the small size of our dataset. And, as expected, at a given number of trees, setting the maximum features to `None' (i.e., providing each tree with all features) was found to cause a significant drop in performance due to overfitting.\\

\begin{table}[!t]
\centering
\begin{adjustbox}{width=\textwidth}
\begin{tabular}{@{}ccccc@{}}
\toprule
\multirow{2}{*}{Rank} & \multicolumn{4}{c}{Feature} \\ \cmidrule(l){2-5} 
  & \textit{PP}                      & \textit{AS}                      & \textit{AHK}                            & All    \\ \midrule
1 & Bass onset strength     & Bass onset strength     & Bass onset strength            & Bass onset strength    \\
2 & Treble energy sum       & MFCC 1                  & MFCC 12                        & Treble energy sum      \\
3 & Treble onset strength   & MFCC 4                  & Treble energy sum              & MFCC 1                 \\
4 & MFCC 3                  & MFCC 3                  & Treble early decay rate        & Treble onset strength  \\
5 & Treble early decay rate & Treble energy sum       & Log attack time                & Treble early decay rate\\
6 & MFCC 1                  & Spec. centroid mean     & \textbf{Spec. centroid stdev.} & MFCC 2                \\
7 & MFCC 2                  & Treble onset strength   & Spec. centroid mean            & MFCC 3                \\
8 & ZCR mean                & Treble early decay rate & MFCC 2                         & Spec. centroid mean   \\
9 & Log attack time         & ZCR mean                & \textbf{Bass early decay rate} & Log attack time       \\
10& MFCC 4                  & MFCC 12                 & ZCR mean                       & ZCR mean              \\ \bottomrule
\end{tabular}
\end{adjustbox}
\caption{The 10 highest ranked features based on importance by a random forest model when fit on different training subsets (entries in bold are features important only to one subset).}
\label{tab:feat_importances}
\end{table}

\begin{figure}[!t]
     \centering
     \begin{subfigure}[b]{\textwidth}
         \centering
         \includegraphics[width=0.6\textwidth]{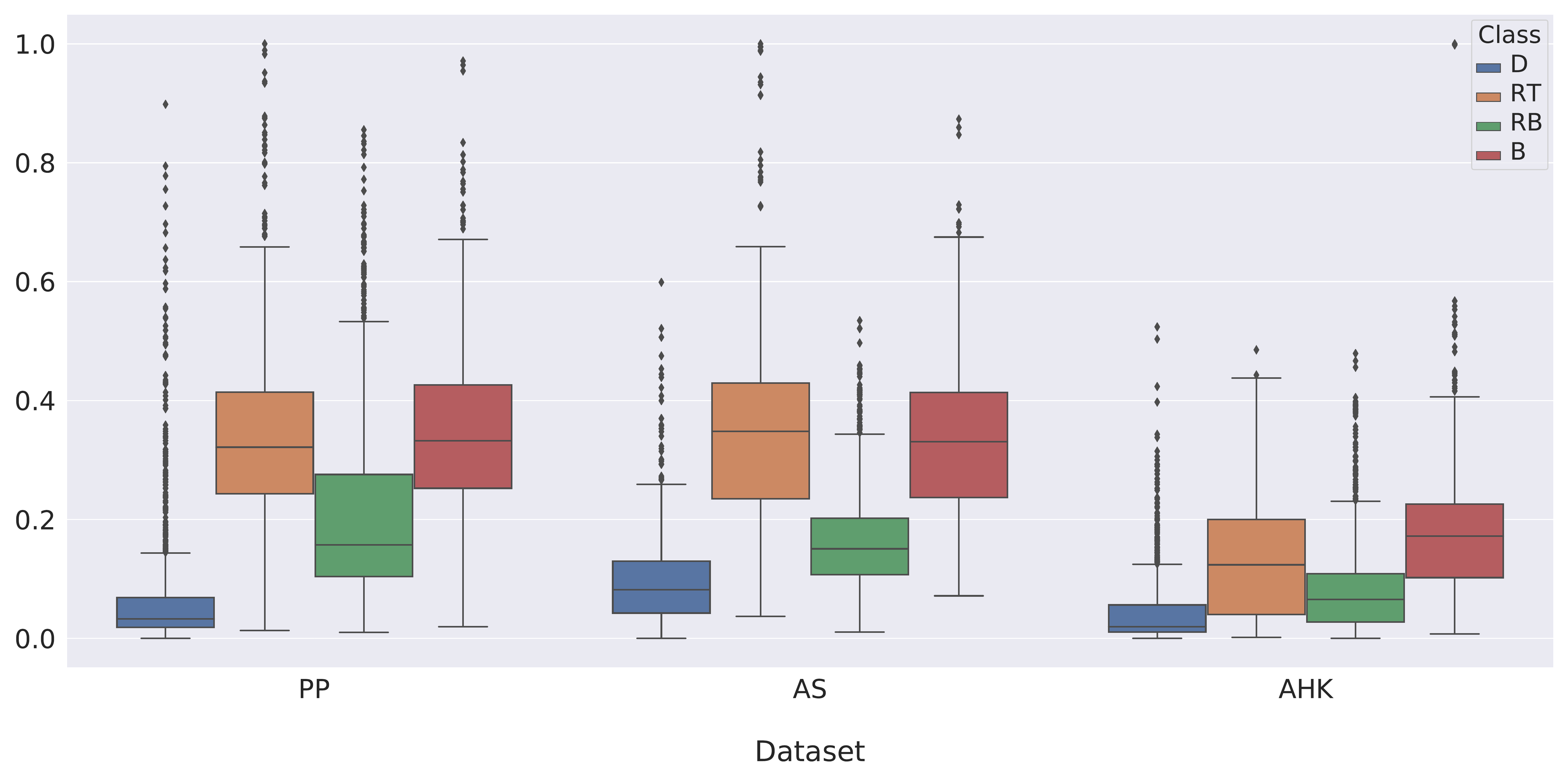}
         \caption{Treble energy sum}
         \label{fig:feat_dist_treble_ener_sum}
     \end{subfigure}
     
     \begin{subfigure}[b]{\textwidth}
         \centering
         \includegraphics[width=0.6\textwidth]{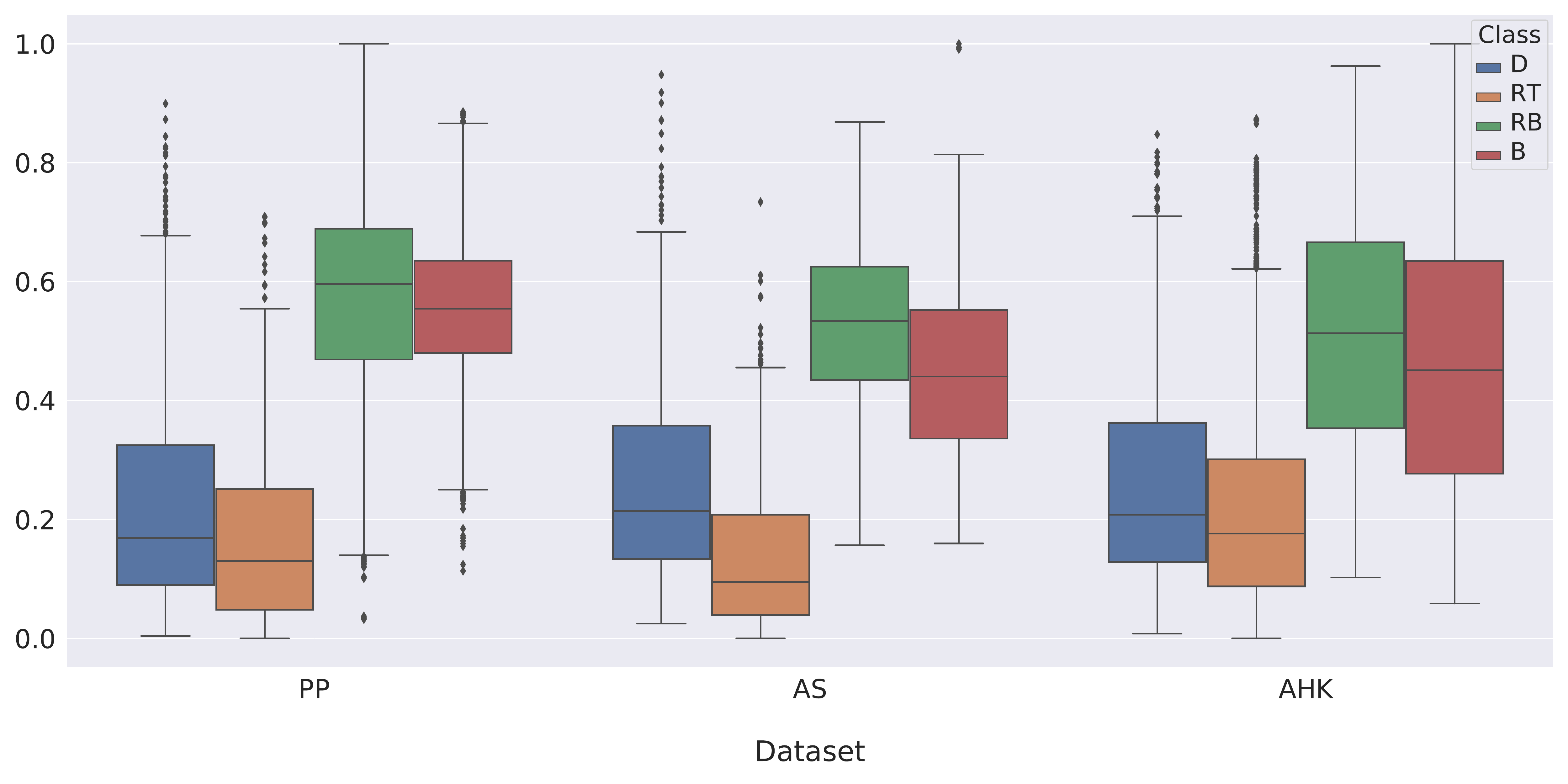}
         \caption{Bass onset strength}
         \label{fig:feat_dist_bass_onset_str}
     \end{subfigure}     
        \caption{Boxplots of the distributions in each training subset of (a) treble energy sum and (b) bass onset strength. (D: `Damped', RT: `Resonant Treble', RB: `Resonant Bass', B: `Resonant Both')}
        \label{fig:feat_dist}
\end{figure}

Finally, given the lower performance on the \textit{AHK} set in the LOTO-CV, a closer look at it can help analyse the instrument dependence of the task. As a first step, we look at models fit separately on each of the three subsets in the training set, with each subset augmented using the pitch-shifting method. Table \ref{tab:feat_importances} shows the ten most important features according to their contribution to the model fit on each tabla set, in comparison with a model fit on the entire training set. The features highlighted in bold refer to those that are exclusive to a subset. We find that almost all the features are common to the three subsets, although with slightly different rankings. This can be taken as a point in favour of the aptness of the features for the stroke classification task, across tabla sets. However, the feature distributions shown in Figure \ref{fig:feat_dist}, of the top two features (from the last column in Table \ref{tab:feat_importances}) shed light on the instrument dependence of the feature values. While the distribution of bass onset strength is similar in all the subsets, in the case of treble energy sum, the \textit{PP} and \textit{AS} subsets have a similar distribution that is quite different from that of \textit{AHK}. This is a likely cause of the poor generalisation of the model.\\

\section{Concluding Remarks}
The presented study brings out the challenges in transcribing tabla stroke sequences when labeled training data from the same tabla set is unavailable. Further, the imbalanced nature of the datasets built using recordings of tabla compositions makes it difficult to train an unbiased model, making it necessary to employ appropriate data balancing and augmentation strategies. The use of pitch-shifting augmentation to increase the size of the less populated resonant stroke categories is found to help, but not on all tabla sets - the cross-validation scores do not improve, while the test set scores do. On the other hand, augmenting in the feature space by interpolating between feature vectors improves the cross-validation scores but leads to an overfit when applied to the entire training set, thus not really improving the test set scores. 
A preliminary analysis of the feature importances in the models trained separately on each tabla set reveals that the same few features are ranked highest in each case. Most of these features are not the standard MFCC or spectral shape-related ones, but happen to be ones that are easier to relate to tabla acoustics (such as onset strength, decay rate, etc). However, the feature distributions vary significantly across tabla sets consistent with the observed instrument-dependence. Therefore, a way to perform more effective data augmentation would be to selectively transform more features that represent the tabla instrument characteristics. The augmentation methods used in the present work are general and do not leverage any specificity of the task. 
Future work will target acquiring a larger dataset, possibly using unsupervised labeling to reduce the manual transcription effort. Further, data augmentation methods that modify those acoustic attributes that are not important to the distinctions between strokes will be explored to better embed the necessary diversity in the training dataset. Finally, we look forward to a system that can be coupled with musical source separation to achieve the automatic transcription of tabla accompaniment in the typical context of a vocal concert's mixed audio.

\pagebreak
\printbibliography

\end{document}